# Revisiting the WMAP – NVSS angular cross correlation.
# A skeptic's view.

Carlos Hernández–Monteagudo

Max Planck Institut für Astrophysik, Karl Schwarzschild Str.1, D-85741, Garching bei München, Germany  
e-mail: `chm@mpa-garching.mpg.de`

**ABSTRACT**

In the context of the study of the Integrated Sachs Wolfe Effect (ISW), we revisit the angular cross correlation of WMAP Cosmic Microwave Background (CMB) data with the NRAO Very Large Array Sky Survey (NVSS). We compute 2-point cross functions between the two surveys, both in real and in Fourier space, paying particular attention on the dependence of results on the flux of NVSS radio sources, the angular scales where correlations arise and the comparison with theoretical expectations. We reproduce previous results that claim an excess of correlation in the angular correlation function (ACF), and we also find some (low significance, ∼ 2–$\sigma$) similarity between the CMB and radio galaxy data in the multipole range $l \in [10, 25]$. However, the signal to noise ratio (S/N) in the ACFs increases with higher flux thresholds for NVSS sources, but drops a ∼ 30 − 50% in separations of the order of a pixel size, suggesting some residual point source contribution. When restricting our analyses to multipoles $l < 60$, we fail to find any evidence for cross correlation in the range $l \in [2, 10]$, where according to the model predictions and our simulations ∼ 50% of the S/N is supposed to arise. Also, the accumulated S/N for $l < 60$ is below 1, far from the theoretical expectation of S/N∼ 5. Part of this disagreement may be caused by an inaccurate modeling of the NVSS source population: as in previous works, we find a level of large scale ($l < 70$) clustering in the NVSS catalog that seems incompatible with a high redshift population. This large scale clustering excess is unlikely to be caused by contaminants or systematics, since it is independent of flux threshold, and hence present for the brightest, most clearly detected ($> 30\sigma$) NVSS sources. Either our NVSS catalogs are not probing the high redshift, large scale gravitational potential wells, or there is a clear mismatch between the ISW component present in WMAP data and theoretical expectations.

## 1. Introduction

For more than ten years there has been observational evidence that the Universe is undergoing a phase of accelerated expansion. Initially motivated by the study of light curves of Super Novae of type Ia at high redshift, this scenario has been supported further by the outcome of subsequent large scale cosmological surveys, like the 2dF survey (Cole et al. 2005), the Sloan Digital Sky Survey[1] (SDSS, Eisenstein et al. 2005) or the *Wilkinson Microwave Anisotropy Probe*[2] [WMAP]. One direct consequence of this accelerating phase is that the growth of density and velocity perturbations is modified, and that the large scale gravitational potentials, which would remain constant in the absence of acceleration, become shallower.

Photons of the Cosmic Microwave Background (CMB) crossing those potential wells should gain energy, since they would leave a potential well that is shallower than at the time of entering it. This late gravitational blue shift on the CMB photons is known as the Integrated (or late time) Sachs Wolfe effect, (hereafter ISW). In the standard WMAP LCDM cosmological scenario, this effect occurs at relatively low redshifts ($z \in [0.5, 1.1]$), and since it involves the non-local, nearby, large scale gravitational potentials, it projects CMB temperature or intensity fluctuations on the large angular scales.

Moreover, as first noted by Crittenden & Turok (1996), those potential wells are regions where structure grows faster and halos preferentially form. If those halos host galaxies, then the gravitational potential wells may be traced by galaxy surveys, and the ISW may be detected after cross correlating the galaxy distribution (in the relevant redshift range) with CMB observations.

After the initial attempts of Boughn & Crittenden (2002) on COBE CMB data, the first detection claims appeared shortly after the first data releases of WMAP, (Nolta et al. 2004; Boughn & Crittenden 2004; Fosalba et al. 2003; Fosalba & Gaztañaga 2004; Vielva et al. 2006; Cabré et al. 2006). Those works computed 2-point cross functions between different galaxy surveys and WMAP temperature maps, both in real and in Fourier space. All those detections were at low significance (below 4–$\sigma$), and only more recent works combining different Large Scale Structure surveys yield detections at the 4,5 –$\sigma$ level, (Giannantonio et al. 2008; Ho et al. 2008). One must remark, however, there is a number of more recent works disputing the results on the SDSS – WMAP cross correlation, claiming that there is no statistical evidence for ISW when looking at the cross two point function between those surveys, (Bielby et al. 2010; López-Corredoira et al. 2010; Sawangwit et al. 2010). Those results are based upon the computation of the error bars of the cross two point function via different methods (Monte Carlo simulations, map rotation, bootstraping, etc), finding that the uncertainty in the measured cross function was too large to assign any statistical significance to it. It must be noted that the signal to noise ratio expected for an ISW cross correlation between a SDSS-like survey with WMAP data is rather small (< 1.5), see Hernández-Monteagudo (2008). Among all surveys used, the NRAO Very Large Array Sky Survey (NVSS) seems to be the one providing the highest significance of the ISW detection claims. The ISW not only shows up at low multipoles (or big angles, demanding a large sky coverage), but also requires deep galaxy surveys sampling the relevant redshift range. These two requirements make the NRAO survey ideal, since it provides a catalog of extragalactic radio sources above the ecliptic latitude of $b_E = -40°$, and supposedly probes the high redshift radio source population. One must note, however, that different claims of ISW detection based upon this galaxy survey are not necessarily consistent to each other: as we shall show below, different works place the detection of the ISW - NVSS cross correlation at different angular scales, under different significance levels, after finding (and correcting for) disparate systematics in NVSS data.

---
[1] SDSS URL site: `http://www.sdss.org`  
[2] WMAP's URL site: `http://map.gsfc.nasa.gov`

This may be of relevance, since the ISW is a *linear* effect which can be, *a priori*, accurately predicted provided that the redshift dependence of the bias and the source number density is well characterized. Actually, as shown in Hernández-Monteagudo (2008) (hereafter Paper I), even a poor description of the redshift distribution of sources should not prevent from making clear predictions on the angular scales where the ISW – galaxy cross correlation should be detected (at least under WMAP cosmogony). Moreover, non linear effects should be of negligible significance in the ISW – density correlation (Smith et al. 2009), although, as Schaefer et al. (2009) demonstrated, imposing constraints on cosmological parameters from ISW observations is indeed sensitive to the accuracy of the source bias characterization.

In this work we revisit the WMAP – NVSS cross correlation in the context of the search for the ISW. The papers is organized as follows: in Section 2 we briefly describe the ISW effect and how its cross correlation with the density field theoretically arises, revisiting in detail the angular/multipole range of relevance in terms of the S/N. In this context we outline the different cross correlation algorithms used in the paper. In Section 3 we describe the NVSS survey, the flux cuts we apply and the interpretation of its auto power spectrum, while in Section 4 we describe the CMB data from WMAP. Section 5 discusses the output of our cross correlation techniques, first on ideal mock CMB and galaxy templates, and then on real WMAP and NVSS catalogs. Results are discussed in Section 6, and final conclusions are presented in Section 7. Throughout this paper, we shall adopt the following set of cosmological parameters, which are motivated by WMAP observations: $\Omega_b = 0.0462$, $\Omega_\Lambda = 0.7208$, $\Omega_m = 0.233$, $n_s = 0.96$, $h = 0.71$ and $\tau_T = 0.084$.

This work is partially motivated by the results of Hernández-Monteagudo et al. (2006), where a matched filter approach in real space yielded no aparent evidence of ISW when cross correlating WMAP data with 2MASS, SDSS and NVSS galaxy surveys.

## 2. The ISW – LSS cross correlation

In this section we describe theoretically how the correlation between ISW anisotropies and projected density fluctuations arises. We conduct this description in Fourier (or multipole) and Real space separately. In each case, we introduce the statistical methods that we shall apply when searching for the ISW – density correlation from both mock simulations and real data.

The spatial fluctuations in the gravitational potential arise on scales typically larger that those of the density fluctuations, due to the $k^{-2}$ factor introduced by the Poisson equation

$$\phi_{\mathbf{k}} = -4\pi G \rho_{b,0} \frac{\delta_{\mathbf{k}}}{a\,k^2}. \tag{1}$$

In this equation $\phi_{\mathbf{k}}$ is the Fourier transform of the gravitational potential field and $\delta_{\mathbf{k}}$ the Fourier mode of the density constrast $\delta(\mathbf{x}) \equiv (\rho(\mathbf{x}) - \rho_b)/\rho_b$, while $\rho_{b,0}$ is the value of the background matter density $\rho_b$ at present. Units for $\rho_b$ and $\mathbf{k}$ are comoving, $G$ is the Newton's gravitational constant and $a$ is the scale factor of the universe. The CMB anisotropies are not sensitive to the linear potential field, but to its time derivative (which causes a net change in the energy of the CMB photons, regardless of their frequency). This gravitational blue/redshift is what is known as the ISW effect, (Sachs & Wolfe 1967)). According to the concordance model, it is only recently when gravitational potentials must have become shallower due to the accelerated expansion of the universe, ($z < 2$). This means that only potential fluctuations below that redshift are projected on the observer (in the form of ISW temperature anisotropies). If a temperature field on the celestial sphere is decomposed on a basis of of spherical harmonics,

$$\frac{\Delta T}{T_0}(\hat{\mathbf{n}}) = \sum a_{l,m} Y_{l,m}(\hat{\mathbf{n}}), \tag{2}$$

then the multipole coefficients corresponding for the ISW temperature field read as, (e.g., Cooray 2002),

$$a_{l,m}^{ISW} = (-i)^l (4\pi) \int \frac{d\mathbf{k}}{(2\pi)^3}\, Y_{l,m}^\star(\hat{\mathbf{k}}) \times$$
$$\int dr\, j_l(kr) \frac{-3\Omega_m H_0^2}{k^2}\,\frac{d(D(r)/a)}{dr} \delta_{\mathbf{k}}. \tag{3}$$

In this equation, $H_0$ is the Hubble parameter, $\Omega_m$ is the matter density parameter and $D(r)$ is the matter linear growth factor. The symbol $j_l(x)$ denotes the spherical Bessel function of order $l$. In the same way, if a galaxy survey samples the density field in a given range of distances (or redshifts) according to some window function $W_g(r)$, then the projected galaxy density field will have these multipole coefficients (Hernández-Monteagudo 2008):

$$a_{l,m}^g = (-i)^l (4\pi) \int \frac{d\mathbf{k}}{(2\pi)^3}\, Y_{l,m}^\star(\hat{\mathbf{k}}) \times$$
$$\int dr\, j_l(kr)\, W_g(r)\, r^2\, n_g(r) b(r,k)\, D(r)\, \delta_{\mathbf{k}}. \tag{4}$$

The symbol $b(r,k)$ denotes the time and scale dependent bias by which galaxies (whose average number density is given by $n_g(r)$) trace the matter distribution. In both equations, we find an integral along the line of sight (LOS) of fields (either potentials or densities) that are connected to the initial scalar metric perturbation field. As first noted by Crittenden & Turok (1996), this introduces an intrinsic correlation between the ISW temperature anisotropy field and a galaxy catalog sampling the accelerated universe. This cross-correlation can be measured either in real (angle) or Fourier (multipole) space. We next outline the methods that we have considered.

### 2.1. In Fourier space

As shown in Paper I, this seems to be the most natural space where conducting the cross-correlation analyses. It is in this space where the linear theory makes its predictions for the ISW, the projected galaxy density and the cross-correlation of both. Furthermore, under full sky coverage ($f_{sky} = 1$) and linear theory, multipole coefficients for different $l, m$ should be independent. Due to the homogeneity and isotropy intrinsic to the theory, there is, a priori, no cosmological information on the $m$ index, so observed multipole coefficients are usually averaged on $m$ when comparing observations to theoretical predictions. For instance, an estimate of the all sky cross power spectrum for two signals $X$ and $Y$ at a given multipole $l$ would be given by

$$\langle a_{l,m}^X (\hat{a}_{l,m}^Y)^* \rangle = \frac{\sum_{m=-l}^{l} a_{l,m}^X (a_{l,m}^Y)^*}{2l+1}. \tag{5}$$

This should be compared to some theoretical expectation for the multipole $l$, $C_l^{XY}$. Provided that multipole coefficients must be independent for different pairs of $l, m$, then estimates of $C_l^{XY}$ (as the one given by Equation (5)) for different $l$ multipoles should also be independent. Hence, in this space it is possible to assign information to different angular (or multipole) scales *independently*. This constitutes a fundamental difference with respect to estimators defined in real space. Of course, if there is a sky mask and only a fraction of the sky is available ($f_{sky} < 1$), then nearby multipole coefficients become partially correlated. Nevertheless this can be accounted for, and in all practical situations it is possible to find a set of multipoles $l$-s in which multipole coefficients remain effectively independent.

Let us now take our $X$ field above to be CMB anisotropy field, and $Y$ a projected galaxy density field obtained under a redshift window function identical to that given by Ho et al. (2008) for the NVSS radio galaxies. This model is explicited by their Equation (33), and covers the relevant redshift range for ISW studies. In what follows, we shall use this window function for our density tracer unless otherwise explicited, although a formal description of this redshift selection function will be outlined in Section 3. The left panel in Figure (1) shows the theoretical expectations of the auto and cross power spectra for these two fields. Since both the ISW and the projected density field are integrals of the matter density contrast (Equations 3, 4), its cross power spectrum (dot-dashed line) is given by (e.g., Cooray 2002; Hernández-Monteagudo 2008)

$$C_l^{ISW,g} = \left(\frac{2}{\pi}\right) \int k^2 dk \, P_m(k) \int dr_1 \, j_l(kr_1) \frac{-3\Omega_m H_0^2}{k^2} \frac{d(D/a)}{dr_1} \times$$

$$\int dr_2 \, j_l(kr_2) \, r_2^2 \, W_g(r_2) \, n_g(r_2) b(r_2, k) \, D(r_2), \tag{6}$$

where $P_m(k)$ denotes the matter power spectrum. This is nothing but the theoretical expectation for the correlation $\langle a_{l,m}^{ISW} (a_{l,m}^g)^* \rangle$ to be measured from our maps (à la Equation 5). The cross power spectrum peaks at around $l \sim 50 - 80$, and at $l \sim 100$ and $l \sim 2$ shows very similar amplitudes. However, as shown in Paper I, the figure of merit is actually the amount of correlation signal at each multipole $l$,

$$\left(\frac{S}{N}\right)_l^2 = \frac{f_{sky} \left(C_l^{ISW,g}\right)^2 (l+1)^2}{\left[C_l^{CMB} C_l^g + (C_l^{ISW,g})^2\right] (l/2+1)}. \tag{7}$$

This is shown in the middle panel of Figure (1), and the cumulative amount of S/N below a given multipole, defined as

$$s_l \equiv \sqrt{\frac{\sum_{l'=2}^{l'=l} (S/N)_{l'}^2}{\sum_{l'=2}^{l'=l_{max}} (S/N)_{l'}^2}}, \tag{8}$$

is given by the right panel: about *half* of the total S/N should be found at $l < 10$, and about 90% of the total S/N should be below $l = 40$. It was shown in Paper I that the actual shape of the cumulative S/N is largely un-sensitive to the particular redshift distribution of the galaxies, or variations of the cosmological parameters around the preferred values of WMAP cosmology. Two different variations from the reference model are shown in the right panel: *(i)* the case where $\Omega_\Lambda = 0.8$ (dashed line) and *(ii)* the case where the galaxy number versus redshift ($dn_g/dz$) is uniform in the redshift range $z \in [1.25, 3]$, and zero otherwise (dot-dashed line). Changes of the cumulative S/N with respect our reference model are unimportant: even an extreme difference in the redshift distribution (as the one depicted by the dot-dashed line) shift the multipole for cumulative S/N of 0.5 from $l = 10$ to $l \simeq 14$, and from $l = 40$ to $l \simeq 53$ in the case of $s_l = 0.9$. Afshordi (2004) shows similar results in terms of the angular/multipole distribution of the S/N. Had we looked at the ratio $(S/N)^2$, then its cumulative value would have reached the 50% of its total value at $l \simeq 18$. We prefer however to use the cumulative S/N rather than $(S/N)^2$ since measurements are quoted at a given S/N level. The reader must note, however, that both quantities are different and that, due to the square root present in Equation (8), $s_{l_{max}} - s_{l=10} \neq 0.5$ even if $s_{l=10} = 0.5$, where $l_{max}$ is arbitrarily large.

In the Fourier part of our analysis, we consider the two methods based in multipole space that were also used in Paper I, namely the Angular Cross Power Spectrum (ACPS) and the Matched Filter (MF). Let us adopt the following model for the data under analysis: we first decompose the measured CMB multipole coefficients $a_{l,m}^{CMB}$ in two contributions, one coming from the Last Scattering Surface ($a_{l,m}^{LSS}$) and another one being the ISW ($a_{l,m}^{ISW}$). At the same time, we consider the multipole coefficients of the galaxy (density) survey ($a_{l,m}^g$). In our LCDM assumed cosmology, the ISW is correlated to the density and hence to the galaxy distribution. Both fields are approximated as Gaussian, so their cross-correlation can be described by breaking the ISW in two terms,

$$a_{l,m}^{ISW} = \alpha_l a_{l,m}^g + \gamma_{l,m}, \tag{9}$$

where the last one ($\gamma_{l,m}$) is uncorrelated with the galaxy field $a_{l,m}^g$, and $\alpha_l = C_l^{ISW,g}/C_l^{gg}$, (e.g., Cabré et al. 2007). We study the correlation between the galaxy survey and the ISW part of the CMB data in separate angular scales, defined by bins containing all multipole coefficients $a_{l,m}$'s verifying $l \in [l_{min}^i, l_{max}^i]$ for the $i$-th bin.

The ACPS computes, for each bin, the statistic $\alpha_i^{ACPS}$:

$$\alpha_i^{ACPS} = \frac{\sum_{l=l_{min}^i}^{l_{max}^i} \sum_m a_{l,m}^{CMB} (a_{l,m}^g)^*}{\sum_{l=l_{min}^i}^{l_{max}^i} \sum_m a_{l,m}^g (a_{l,m}^g)^*}, \tag{10}$$

(with the superscript $*$ denoting 'complex conjugate'), while the MF is sligthly more sophisticated:

$$\alpha_i^{MF} = \frac{\mathbf{t}_i^t \mathbf{C}_{i,i}^{-1} \mathbf{m}_i}{\mathbf{m}_i^t \mathbf{C}_{i,i}^{-1} \mathbf{m}_i}. \tag{11}$$

The array $\mathbf{t}_i$ refers to the CMB multipole coefficients (containing both the LSS and the ISW contribution) in the $i$-th multipole bin, whereas the array $\mathbf{m}_i$ contains the galaxy survey multipole coefficients for the same bin. We adopt the same notation as in Paper I, for which given an array $\mathbf{a}$, $(\mathbf{a})_j$ denotes the $j$-th component of that array, but $\mathbf{a}_j$ denotes the $j$-th array of a general set of arrays $[\mathbf{a}_1, \mathbf{a}_2, ..., \mathbf{a}_n]$. The covariance matrix $\mathbf{C}_{i,i}$ is *a priori* defined as

$$\mathbf{C}_{i,i} = \langle (\mathbf{t}_i^t - \alpha_i^t \mathbf{m}^t)(\mathbf{t}_i - \alpha_i \mathbf{m})^* \rangle. \tag{12}$$

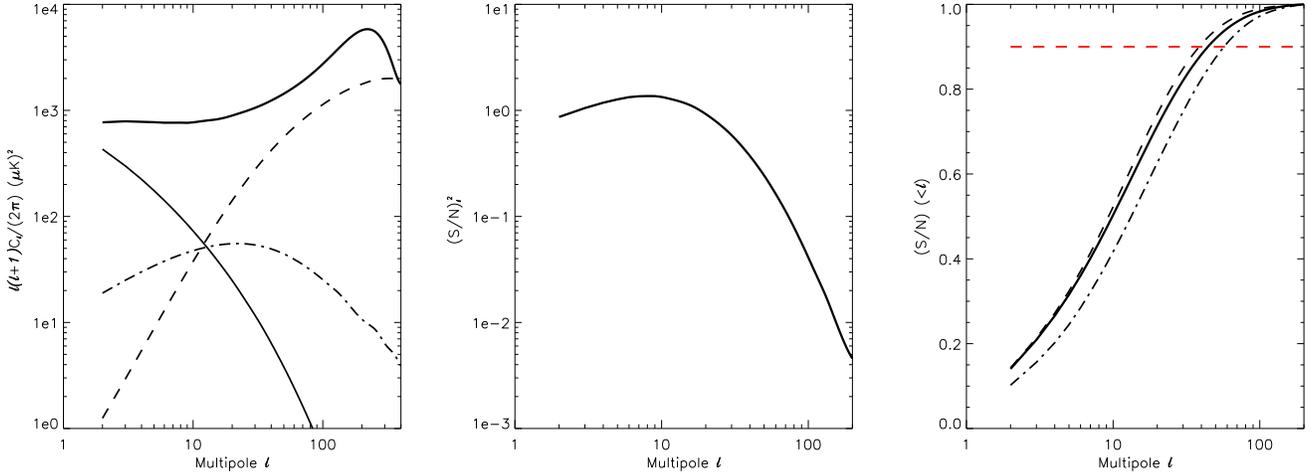

**Fig. 1.** *Left panel:* Auto CMB power spectrum (thick solid line), auto ISW power spectrum (thin solid line), auto galaxy power spectrum (dashed line) and ISW – galaxy cross power spectrum (dot-dashed line). The latter two have arbitrary units, for display purposes. *Middle panel:* S/N corresponding to the ISW – galaxy cross power spectrum per multipole $l$. *Right panel:* Cumulative S/N of the ISW – galaxy cross power spectrum below a given multipole $l$.

That is, it accounts for all components in **t** that are *not* correlated to **m**, namely the sum $a_{l,m}^{LSS} + \gamma_{l,m}$. However, in this work we shall consider only the part generated at the LSS, and ignore the part of the ISW that is not correlated to the galaxy survey ($\gamma_{l,m}$). If we had a precise characterization of the redshift distribution of our galaxy survey, then it would make sense to compute $\mathbf{C}_{i,i}$ exactly, as shown in Frommert et al. (2008). For our purposes, considering only the LSS component when computing the covariance matrix shrinks our errors on the cross-correlation analyses. This means that if our analyses yield no significant correlation between the CMB and the galaxy survey, these must be regarded as *conservative*: a more precise approach when computing the covariance matrix would assign more statistical significance to the apparent lack of correlation. This approach is adopted also for the ACPS: when running Monte Carlo simulations to estimate the dispersion of $\alpha_i^{ACPS}$, we shall consider only the LSS component. When quoting S/N ratios to the significance of the cross-correlation, we shall use the $\beta$ statistic for both MF and ACPS methods. This statistic is defined as:

$$\beta_j \equiv \sum_{i=1}^{j} \frac{\alpha_i}{\sigma_{\alpha_i}}, \quad (13)$$

where the rms dispersion on the $\alpha_i$-s are computed from 10,000 MC simulations (as outlined in Paper I). The $\beta_j$ statistic is Gaussian distributed provided the $\alpha_i$-s are Gaussian distributed (as it is found when looking at the MC simulations), and the S/N up to a given multipole bin $j$ is then given by

$$\left(\frac{S}{N}\right)_j = \frac{\beta_j}{\sigma_{\beta_j}}, \quad (14)$$

where again the rms of the $\beta_j$-s was found from the 10,000 MC simulations. Such estimate of $\sigma_{\beta_j}$ should account for the existing correlations among different $\alpha_i$ estimates whenever $f_{sky} < 1$. Note that the $\beta_j$ statistic is sensitive to an integrated excess of correlation *or* anticorrelation present in the multipole interval ranging from $i = 1$ to $i = j$. In the presence of the ISW we expect statistically significant positive values of $\alpha_i$-s in that multipole interval (that is, *correlation*), and hence positive values of $\beta_j$. In the case where the $\alpha_i$'s fluctuate around zero at the level of many $\sigma_{\alpha_i}$-s, the $\beta_j$ statistic might miss their statistical significance, since the sum of ratios in Equation (13) could tend to average out the contribution from each ratio. A simple examination of the individual $\alpha_i$-s should asses whether that is the actual case or not.

We refer to Paper I for more details on the implementation and performance of these two methods. In passing, we remark that the MF tends to be more sensitive under aggressive masks, but both methods perform very similarly for masks close to those of NVSS and WMAP.

### 2.2. In real space

In real space we shall consider the angular cross-correlation function (ACF), defined as

$$w(\theta) = \langle t(\hat{\mathbf{n}}_1) m(\hat{\mathbf{n}}_2) \rangle, \quad (15)$$

where $t$ and $m$ refer to the CMB and galaxy density map, respectively, and $\theta$ equals $\cos^{-1}(\hat{\mathbf{n}}_1 \cdot \hat{\mathbf{n}}_2)$. It is customary to work in real space and compute the ACF as an estimate built upon all pairs of pixels in the *useful* patch of the sky lying a distance $\theta \pm d\theta$ away:

$$w^R(\theta) = \frac{\sum_{i,j \in \theta \pm d\theta} t(\hat{\mathbf{n}}_i) m(\hat{\mathbf{n}}_j)}{\sum_{i,j \in \theta \pm d\theta} 1}, \quad (16)$$

where the superscript $R$ denotes 'computed directly in real space'. When implementing this method, we chose to remove the monopole in both the CMB and density maps *outside* the effective mask, that is, we removed the *mean* of each maps in the common subset of *valid* pixels for *both* surveys[3]. In this way, the product in the numerator of the equation above takes place only between *angular fluctuations*.

Provided that the theory is written preferentially in Fourier space, le us express the ACF in terms of the multipole coefficients for each field:

$$w(\theta) = \sum_{l_1, m_1} \sum_{l_2, m_2} Y_{l_1, m_1}(\hat{\mathbf{n}}_1) Y^*_{l_2, m_2}(\hat{\mathbf{n}}_2) \langle a^{CMB}_{l_1, m_1} (a^g_{l_2, m_2})^* \rangle. \quad (17)$$

---
[3] Note that here we regard the mask as the set of pixels being *excluded* from the analysis.

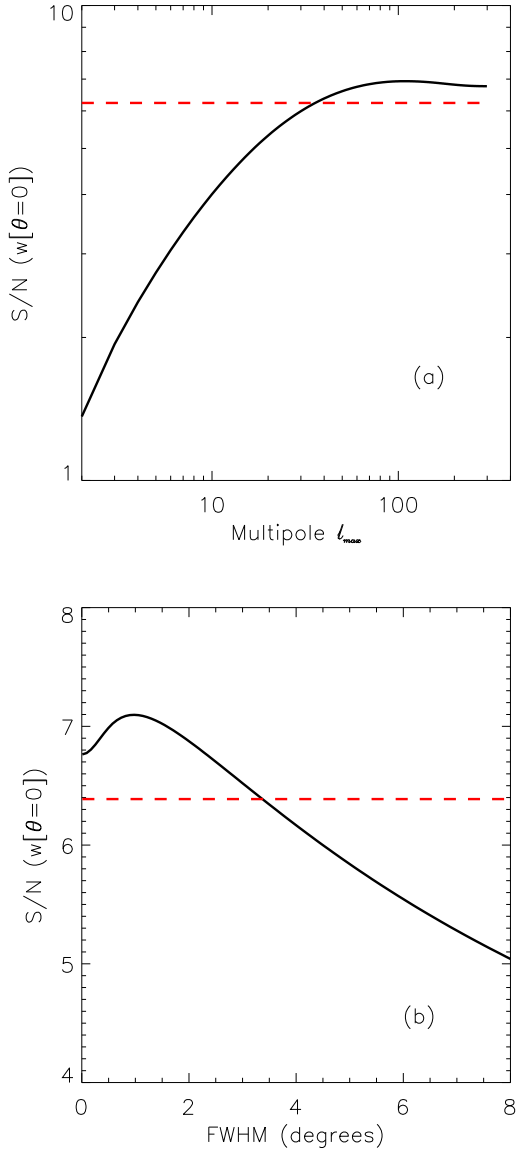

**Fig. 2.** *(a)* S/N of the zero lag angular cross-correlation function (ACF) versus the maximum multipole $l_{max}$ considered for our NVSS reference model (Ho et al. 2008). *(b)* S/N of the ACF versus of the FWHM of the observing Gaussian PSF. The 90% level of the maximum S/N is given by the dashed red line.

If both fields are isotropic, then $\langle a^{CMB}_{l_1,m_1}(a^g_{l_2,m_2})^*\rangle = \delta^K_{l_1,l_2}\delta^K_{m_1,m_2}C^{ISW,g}_l$, i.e., only the diagonal terms contribute, and the last expression can be simplified to

$$w(\theta) = \sum_{l_1} \frac{2l_1+1}{4\pi} C^{ISW,g}_{l_1} P_{l_1}(\hat{\mathbf{n}}_1 \cdot \hat{\mathbf{n}}_2), \qquad (18)$$

where $P_l(\hat{\mathbf{n}}_1 \cdot \hat{\mathbf{n}}_2)$ are Legendre polynomials of the dot product of the angle separating the two lines of sight considered. This expression clearly shows that *all angular* scales (or multipoles *l*-s) are, *a priori*, contributing to the correlation function at a given separation $\theta \equiv \cos^{-1}(\hat{\mathbf{n}}_1 \cdot \hat{\mathbf{n}}_2)$. If some excess of cross-correlation is found using the ACF, it should be found at low $\theta$'s. This follows from the fact that, if two fields are correlated in Fourier/multipole space and have positive cross-spectra (as it is the case for the $C^{ISW,g}_l$-s), then the contribution from all multipoles of these cross spectra is going to have maximum contribution at zero lag ($\theta = 0$) provided that $P_l(1) = 1$ for all *l*-s, (see Equation 18). For $\theta > 0$, Legendre polynomia of order $l \sim 1/\theta$ start to oscillate around zero, dumping the contribution of the corresponding $C^{ISW,g}_l$-s. However, at zero lag, it will not be possible to distinguish the particular angular scale (or multipole *l*) assigned to the signal giving rise to the cross-correlation excess. From Equation (18) it is easy to see as well that estimates of the ACF at different angular bins will be highly correlated: the *same* whole set of cross power spectra $C^{ISW,g}_l$'s will be contributing at different $\theta$'s, and each contribution will be modulated by the amplitude of the corresponding Legendre polynomial $P_l(\hat{\mathbf{n}}_1 \cdot \hat{\mathbf{n}}_2)$. How could one distinguish the presence of an ISW-induced cross correlation from other sources of cross correlation? Let us make specific predictions for the ISW – density correlation. We shall next focus our analyses at the zero lag ($\theta = 0$) point. Let us assume that our *fixed* galaxy survey is deep enough so that it is correlated to the whole ISW temperature field. In this situation, the source of uncertainty is only the part of the CMB being generated at the Last Scattering Surface. In this case, it is easy to prove that the S/N associated to the ACF reads as

$$\left(\frac{S}{N}\right)[w(\theta=0)] = \frac{\sum_{l=2}^{l_{max}}(2l+1)C^{ISW,g}_l(W^{PSF}_l)^2 / (4\pi)}{\sqrt{\sum_{l=2}^{l_{max}}(2l+1)C^g_l C^{LSS}_l (W^{PSF}_l)^4 / (4\pi)^2}}$$

$$= \frac{\sum_{l=2}^{l_{max}}(2l+1)C^{ISW,g}_l(W^{PSF}_l)^2}{\sqrt{\sum_{l=2}^{l_{max}}(2l+1)C^g_l C^{LSS}_l (W^{PSF}_l)^4}}. \qquad (19)$$

This equation assumes that $f_{sky} = 1$ and we have explicitely included the window function of the (spherically symmetric) PSF/beam of the experiment ($W^{PSF}_l$), by which the observed Fourier multipole of a field on the sphere is given by $a^{obs}_{l,m} = W^{PSF}_l a_{l,m}$, with $a_{l,m}$ the real underlying multipole and $a^{obs}_{l,m}$ the observed one. In Figure (2a) we display the S/N for the ACF versus the maximum multipole $l_{max}$ present in Equation (19). In this case, we assume a perfect PSF (so $W_l = 1 \forall l$). The S/N shows a maximum at around $l \sim 80$, although it is rather flat in a wide range of multipoles. A real space analogue of this plot is provided in Figure (2b), which displays the dependence of the S/N of the ACF versus the FWHM of the Gaussian PSF of the observing instrument. For a Gaussian PSF, we have that $W^{PSF}_l = \exp(-l(l+1)\sigma^2/2)$ and $\sigma = \text{FWHM}/\sqrt{8\log 2}$. Even for a PSF as large as FWHM = $4°$, we see that practically 90% of the total S/N ratio is preserved: this confirms that most of the ISW – galaxy cross S/N belongs to the large scales and that the contribution of many other small scale secondary effects may be avoided by the use of a low pass filter. Choices of larger FWHMs (FWHM = 5 - 8 °) should still yield high values of S/N, and this can be used a consistency test for ISW detection claims.

If both (CMB and galaxy template) maps contain some contaminant that is not isotropic, then this will *not only* contribute to the diagonal terms (adding spurious power to the underlying cross power spectrum), but also may introduce (at a given separation $\hat{\mathbf{n}}_1 \cdot \hat{\mathbf{n}}_2$) terms coupling different multipole pairs ($l_1 \neq l_2$, $m_1 \neq m_2$) that are zero in the isotropic case. I.e, if the contami-nant Fourier multipoles are given by $a_{l,m}^{foreg}$, then their impact on Equation (18) can be written as

$$w(\theta) = \sum_{l_1} \frac{2l_1+1}{4\pi} \left[ C_{l_1}^{ISW,g} + C_{l_1}^{foreg,foreg} \right] P_{l_1}(\hat{\mathbf{n}}_1 \cdot \hat{\mathbf{n}}_2) +$$

$$\sum_{l_1,m_1} \sum_{l_2 \neq l_1, \, m_2 \neq m_1} \langle a_{l_1,m_1}^{foreg} (a_{l_2,m_2}^{foreg})^* Y_{l_1,m_1}(\hat{\mathbf{n}}_1) Y_{l_2,m_2}^*(\hat{\mathbf{n}}_2) \rangle, \quad (20)$$

where the last term couples different multipoles, and, *a priori*, is not zero for anisotropic signals. Under non full sky coverage ($f_{sky} < 1$) there is a non-negligible coupling between different *estimated* Fourier multipoles, due to the lack of orthonormality of the spherical harmonics. But, as mentioned above in the context of the inversion of the covariance matrix, this effect can be accounted for when interpreting results.

The last term in Equation (20) can be avoided by constructing an estimator of the ACF which couples only *identical* Fourier multipoles, since this is the signal that the theory predicts. In our study, apart from the standard implementation of Equation (15), we also adopt the following approach:

– We Fourier invert the two maps $t(\hat{\mathbf{n}})$ and $m(\hat{\mathbf{n}})$ (after both being multiplied by a joint mask) and consider the Fourier multipoles $t_{l,m}$-s and $m_{l,m}$-s in separate multipole bins limited by $[l_{min}^i, l_{max}^i]$ for the $l$ index for the $i$-th bin.
– Within each $l$-bin, we invert back onto real space, obtaining a set of maps $t_i(\hat{\mathbf{n}})$ and $m_i(\hat{\mathbf{n}})$, with $i = 1, N_b$ and $N_b$ the number of bins in $l$ considered, (just as for the ACPS and MF methods).
– Finally, at each separation $\theta \equiv \cos^{-1}(\hat{\mathbf{n}}_1 \cdot \hat{\mathbf{n}}_2)$, we considered the matrix of ACFs given by

$$(\mathbf{w})_{i,j}(\theta) \equiv \langle t_i(\hat{\mathbf{n}}_1) m_j(\hat{\mathbf{n}}_2) \rangle, \quad (21)$$

where both $i$ and $j$ run from 1 to $N_b$.

The advantage of this procedure is that, after running Monte Carlo simulations, one can keep track of which multipole bins or angular scales may give rise to some excess signal/cross-correlation. Ideally (i.e., under full sky coverage and isotropic signals), by adding up all elements of the $\mathbf{w}(\theta)$ array one should recover the standard correlation matrix computed *à la* Equation (18). In such sum, off-diagonal terms should have no impact, and the addition of all elements of $\mathbf{w}(\theta)$ should be very close to its trace. In practice, the difference between the standard estimate and the trace, and between the trace and the sum of all elements of $\mathbf{w}(\theta)$ should provide a handle of the impact of the mask and the presence of anisotropic signals. We finish this section by expliciting a couple of technical details. All analyses in Fourier space ignored the contribution from $l = 0, 1$. Also, when inverting the $t_{l,m}$'s, $m_{l,m}$-s back into real space according to our implementation of $\mathbf{w}(\theta)$, we computed this correlation matrix after considering *all* pixels on the sphere (since the effect of the mask was already included when inverting into the multipole coefficients). As mentioned above, when implementing Equation (16) we removed the mean of the maps *outside* the joint (CMB + galaxy survey) mask, and computed the correlation function after considering *only* pixels outside the joint mask. These two approaches are not identical, and yield slight differences that will be commented below.

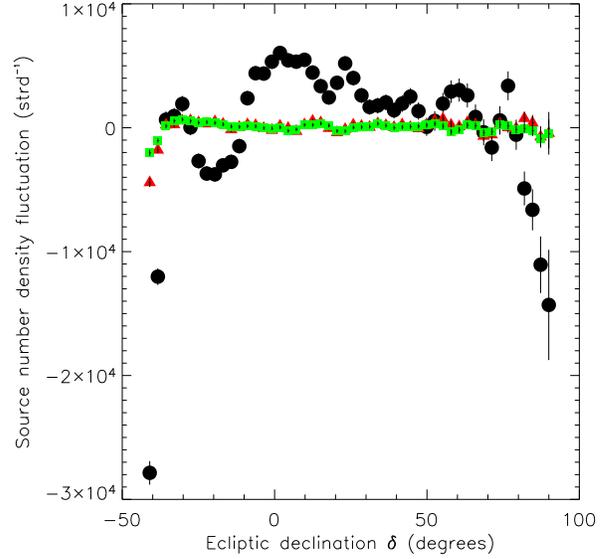

**Fig. 3.** Variation of the NVSS radio galaxy fluctuation versus ecliptic declination for sources brighter than 2.5 mJy (black circles), 30 mJy (red triangles) and 60 mJy (green squares).

## 3. The NRAO Very Large Array Sky Survey [NVSS]

In this section we first describe the main characteristics of the NVSS survey, the main source populations contained, and the flux cuts applied on the data to be analyzed. We next compute the auto power spectrum for each flux cut, paying particular attention on systematic effects associated to the ecliptic declination dependence of the source number density and the sky mask.

### 3.1. Flux cuts in the NVSS survey

Our galaxy survey that we shall use as probe of the gravitational potentials will be the NRAO Very Large Array Sky Survey, (NVSS, Condon et al. 1998). This survey was conducted at a frequency of 1.4 GHz and covered all the sky above ecliptic latitude of $b_E > -40°$. Its galaxy catalog contains around 1.8 million sources, and it is claimed to be 99% complete above a flux limit of 3.5 mJy. Observations were conducted by the Very Large Array (VLA) in two different configurations: the $D$ configuration was used for ecliptic latitudes in the range $b_E \in [-10°, 78°]$, while the $DnC$ configuration was used under large zenith angle ($b_E < -10°$, $b_E > 78°$). As noted by Blake & Wall (2002), this change of configuration introduced some systematics in the galaxy number density. In Figure (3) we plot the fluctuations of the radio galaxy number density (around its mean) versus ecliptic latitude for NVSS sources after considering three different flux thresholds: black circles display the case where the threshold has been imposed at 2.5 mJy, while red triangles and green squares correspond to 30 mJy and 60 mJy, respectively. It is clear that dim sources are strongly affected by the VLA configuration, since the number density fluctuations changes dramatically for the declinations $b_E = -10°, 78°$ where the observing configuration is switched. This does not appreciably happen for the brightest sources (thresholds at 30 and 60 mJy), which show a rather flat pattern versus declination. This is not surprising, since those sources at detected at $> 30-\sigma$ (the average noise level of NVSS is $\sim 0.45$ mJy/beam, Condon et al. (1998)). In addition, Blake & Wall (2002) raise another issue related to dim NVSS sources:

when pointing to a bright radio source, side lobes usually show up surrounding it and being counted as spurious dim sources in the catalog. Although potentially of relevance, this effect should be avoided in the brightest radio sources, since the point source mask built by WMAP team typically cancels a circle of radius 0.6° around the bright radio sources detected by this experiment.

The use of the NVSS in an ISW context is motivated by the fact that luminous Active Galactic Nuclei (AGNs) are supposed to be good tracers of the density field at high redshift. However, among NVSS radio galaxies, one should, *a priori*, distinguish two different source populations, namely high luminosity AGNs and nearby Star Forming Galaxies (SFGs). If the contribution of the latter population is not negligible, then it might *distort* our template of the high redshift density distribution by adding a very low redshift galaxy sample. It was shown in Paper I that, in the concordance model, most of the ISW signal is generated in the redshift range $z \sim [0.5, 1.1]$, and therefore ideally our galaxy survey should probe this epoch. The SFGs are placed at very low redshift ($z < 0.01$) and for this reason provide no information in terms of ISW studies. They are intrinsically less luminous sources in the radio, and, as shown by Condon et al. (1998), dominate the source counts in the low flux end (sub mJy at 1.4 GHz). According to Figure (1) of Condon et al. (1998), they contribute to a ~ 30% of the total number of weighted source counts at 1 mJy, but this contribution should drop rapidly at larger fluxes measured at 1.4 GHz. However, this constitutes another argument to test how correlation tests depend on the flux cut applied to NVSS sources.

In our analyses, we build three different galaxy templates out of NVSS data, each of them corresponding to flux thresholds at 2.5, 30 and 60 mJy, containing $1.61 \times 10^6$, $2.2 \times 10^5$ and $1.1 \times 10^5$ sources respectively. This shows that even under the strictest flux cuts, there are in average several (> 3) sources per square degree, and that Poisson/shot noise should be unimportant for the angular scales of relevance for ISW - density cross-correlation studies. Indeed, we shall find that, on the large scales, cross-correlation analyses will yield very similar results for each of the three catalogs considered here.

### 3.2. The Angular Power Spectrum of the NVSS Survey

In this subsection we study the angular clustering of NVSS sources for different flux thresholds. For a given source population above the *j*-th threshold, it is possible to write the angular power spectrum of the source angular number density as

$$C_l^j = \bar{n}_j^2 \left( C_l^j + \frac{1}{\bar{n}_j} \right), \tag{22}$$

where $\bar{n}_j$ denotes the average source angular density above the *j*-th threshold. The first term in parentheses corresponds to the intrinsic source clustering (which is independent of the source average density), whereas the second one refers to the Poisson/shot noise, and is independent of the multipole $l$. For high flux thresholds we expect having fewer sources, and hence a more important relative weight of the Poissonian shot noise in the angular power spectrum: i.e., for lower values of $\bar{n}_j$, the Poisson term dominates over the clustering term at lower multipoles or larger angular scales. This is shown in the top panel of Figure (4): filled black circles correspond to the angular power spectrum multipoles ($C_l$-s) for the full radio sample above 2.5 mJy. Red triangles and green squares correspond to thresholds $S_\nu > 30, 60$ mJy, respectively. The error bars in the top panel assume a relative uncertainty for each multipole $l$ equal to $\sqrt{2/(2l + 1)/f_{sky}}$, with $f_{sky}$ the un-masked fraction of the sky. This assumes Gaussian statistics ruling the NVSS source distribution on the large scales. If, as suggested recently (Xia et al. 2010), the clustering at large angles is dominated by some non-Gaussian component, the those errors should be revised. On the small scales, the statistics is practically Poissonian, and the scatter with respect to the theoretical prediction is small.

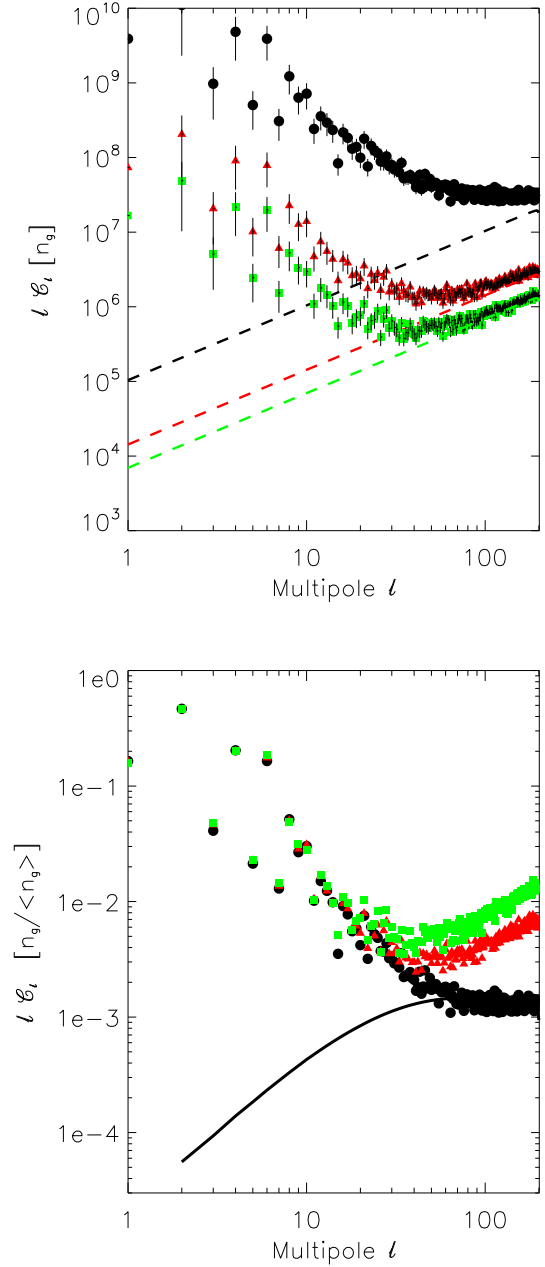

**Fig. 4.** *(Top)* Angular pseudo-power spectrum estimates $C_l$ (times the multipole $l$) for the angular number density of NVSS sources ($n_j(\mathbf{n})$) brighter than 2.5 mJy (black circles, $j = 1$), 30 mJy (red triangles, $j = 2$) and 60 mJy (green squares, $j = 3$). *(Bottom)* Same as top panel but for the normalized angular number density, i.e., for $\delta_j(\mathbf{n}) \equiv n_j(\mathbf{n})/\bar{n}_j$ and $j = 1, 2, 3$. The solid line displays the pseudo-power spectrum of the model of Ho et al. (2008) renormalized to the high $l$ amplitude of the $j = 1$ threshold sample ($S_\nu > 2.5$ mJy).

In all panels we show estimates for the *pseudo-power spectrum*, which are obtained after Fourier inverting the whole celes-tial sphere (even when both NVSS and WMAP masks have been applied). For this we used the HEALPix (Górski et al. 2005) software package. The presence of this effective mask (hereafter denoted as $W(\mathbf{n})$) does have an effect on the estimates of the $C_l$'s: the recovered pseudo power spectrum multipoles are, on average, related to the real underlying power spectrum in the following way (Wandelt et al. 2001):

$$C_l = \sum_{l'} \mathcal{M}_{l,l'} C_{l'}^{full\ sky}. \tag{23}$$

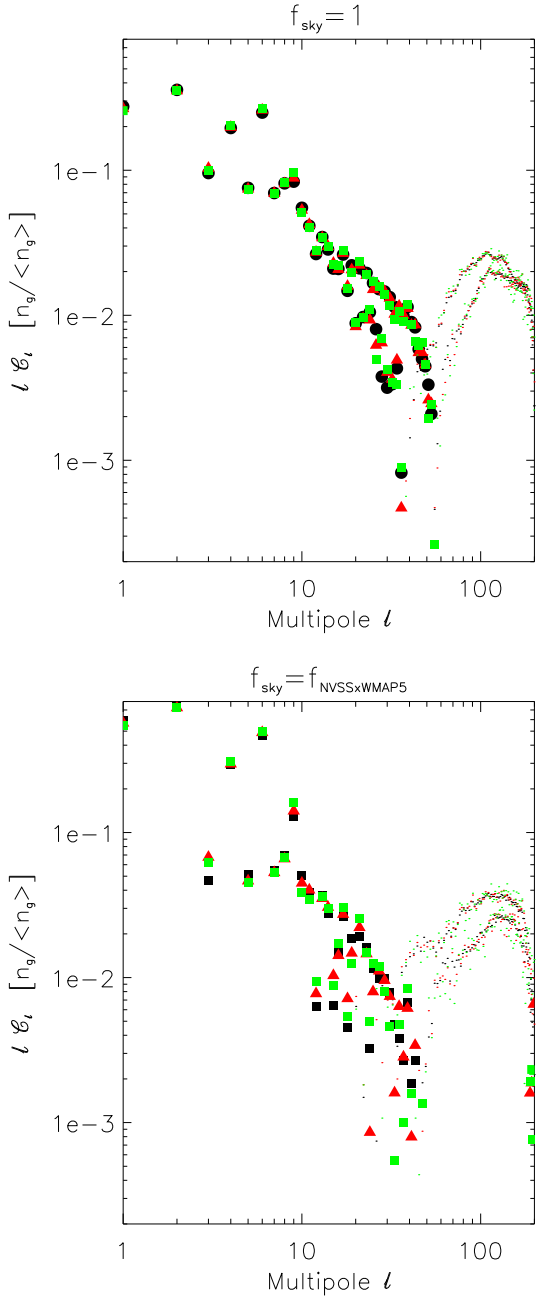

**Fig. 5.** The color coding is as in the previous figure. *(Top)* Angular power spectrum estimates for the *normalized* angular number density of NVSS sources obtained after Legendre inverting the ACF. All pixels on the sphere, including those blank pixels zeroed by the mask, were considered. *(Bottom)* Angular power spectrum estimates from the ACF, but after considering only pixels *outside* the joint mask.

The matrix $\mathcal{M}_{l,l'}$ is defined as

$$\mathcal{M}_{l,l'} = \frac{2l'+1}{4\pi} \sum_{l_3} (2l_3+1) \mathcal{W}_{l_3} \begin{pmatrix} l & l' & l_3 \\ 0 & 0 & 0 \end{pmatrix}^2, \tag{24}$$

where the symbol $\mathcal{W}_l$ denotes the angular power spectrum of the effective mask $W(\mathbf{n})$. I.e., if the mask Fourier multipole $w_{l,m}$ is given by

$$w_{l,m} = \int d\mathbf{n}\ Y_{l,m}^*(\mathbf{n}) W(\mathbf{n}), \tag{25}$$

then its power spectrum can be computed as

$$\mathcal{W}_l = \frac{\sum_{m=-l}^{l} |w_{l,m}|^2}{2l+1}. \tag{26}$$

The array in Equation (24) denotes the 3J-Wigner symbol that vanishes if $l+l'+l_3$ is odd or if $l+l' < l_3$ or $|l-l'| > l_3$. In Figure (4) we are displaying pseudo power spectrum multipoles. Top panel shows the raw pseudo $C_l$'s versus multipole $l$ for each threshold: in each case, we see that they approach the constant Poissonian limit at high $l$ computed out of $\bar{n}_j$ and displayed by the dashed lines. This Poissonian limit has been obtained by first computing the Poisson term for the angular power spectrum ($C_l^{P,j} = \bar{n}_j$) and then converting it into pseudo $C_l$ by means of Equation (23). For higher flux thresholds we find that the Poisson term becomes dominant at lower $l$-s / larger angular scales. The presence of the clustering term at lower $l$-s is also evident. The overall shape of the HEALPix pseudo power spectrum multipoles are in rough agreement with the (already corrected for the mask) $C_l$'s computed by Blake et al. (2004) (see their Figure (7)): they also find a steep decrease of the amplitude of the power spectrum multipoles at low $l$-s, and a *plateau* at smaller angular scales (which in their case is compatible with zero). However, the slope and the amplitudes are different, presumably due to the different flux threshold, multipole binning and mask correction applied in their case.

The intrinsic clustering properties of the radio sources are very similar for sources above different thresholds. This is demonstrated in the bottom panel of Figure (4), where the pseudo power spectrum estimates have been computed after normalizing, in each case, the source angular number density by its average value ($\bar{n}_j$), i.e., after effectively dropping the $\bar{n}_j^2$ prefactor in Equation (22). The coincidence of different symbols at low $l$-s shows that the intrinsic clustering is very similar for sources above different flux thresholds, and in particular, that the clustering of *all* NVSS sources *is very close to the clustering of the brightest, clearly detected NVSS sources.*

The impact of the declination dependence completeness of the radio survey on its power spectrum was assessed by rebuilding our NVSS templates in equatorial coordinates. When computing the pseudo power spectra in this reference frame, we ignored the azymuthal modes ($a_{l,m=0}$ modes) at estimating the averages over $m$, just as in Smith et al. (2007). This operation lowered by more than one order of magnitude the dipole, by a factor of $\sim 2$ the quadrupole, and introduced a correction below the level of $\sim 30\%$ for subsequent multipoles, leaving the pattern of Figure (4) practically unchanged.

In order to quantify the impact of the mask on our pseudo $C_l$ computation, we estimated the NVSS auto power spectrum by Legendre transforming the ACF of each of the NVSS templates. That is, by computing the integral

$$C_l = (2\pi) \int_{-1}^{1} d[\cos\theta]\ w^R(\theta) P_l[\cos\theta], \tag{27}$$

with $P_l(x)$ the $l$-th order Legendre polynomial. As shown by Szapudi et al. (2001), this inversion should, *a priori*, be less sensitive to low $l$ mode coupling introduced by the mask, (although, as shown in Hernández-Monteagudo et al. (2004), at small scalles errors in the $C_l$'s become correlated in a non-trivial way). Due to the choice of the roots of the Legendre polynomials (on which the integrand given in Equation (27) is evaluated), this estimator based on the ACF is *unsensitive* to the Poisson term of the angular power spectrum. Therefore it provides estimates of the clustering $C_l$-s *exclusively*. The top panel of Figure (5) shows the angular power spectra obtained after Legendre inverting the ACF of the three NVSS templates *in the whole sky, i.e., after considering also the masked regions*. The bottom panel of the same figure considered only un-masked pixels. At the multipole range where the clustering term dominates more clearly ($l < 20$), we obtain in both cases values that are very close to those given in the bottom panel of Figure (4). At higher multipoles the clustering term (the only one to which this power spectrum estimator is sensitive) decreases further and becomes dominated by numerical noise and negative, (negative values are displayed by colored dots). We conclude that, for the scales relevant to the clustering, both approaches (pseudo-$C_l$'s and Legendre transform of the ACFs) yield consistent results: a dominant clustering signal is present at low multipoles, being very similar for the three flux cuts under consideration.

When using the full catalog, Ho et al. (2008) also found a high level for the $C_l$'s at low multipoles, but claimed that the low $l$ angular power of NVSS was largely spurious. However, our analyses show that such statement must necessarily apply also to the brightest radio sources detected with S/N greater than $\sim 60$, prompting the question what kind of systematic may bias so severely the clustering properties of so bright and clearly detected sources. On the other hand, Blake et al. (2004) interpret most of the low $l$ NVSS power as being generated at low redshifts ($z < 0.1$), in contradiction with the findings of Ho et al. (2008). We also find that the clustering low $l$ power of NVSS is incompatible with the redshift distribution assigned by Ho et al. (2008) to NVSS sources, as the mismatch between the solid line and the filled circles in the bottom panel of Figure (4) shows. The solid line displays the pseudo auto power spectrum for NVSS sources according to the model of Ho et al. (2008) (after matching to the amplitude of the high $l$ pseudo power spectrum estimates for the $S_\nu > 2.5$ mJy threshold): there is reasonable agreement in the shape of the curve only for $l > 70$, whereas for larger angular scales (which are the ones of relevance in ISW - LSS cross-correlation studies) the presence of excess power is evident. This fact has also prompted other authors (Negrello et al. 2006; Raccanelli et al. 2008) to invoke a decreasing in redshift bias for NVSS sources, which would weight more the low redshift tail of the NVSS source distribution[4]. All these works point to the actual distribution of NVSS sources still being under open debate, with a clear mismatch between their large angle clustering properties and their redshift distribution as suggested by different authors and models.

For the time being, the model of Ho et al. (2008) for NVSS sources will be taken as our reference framework on which we shall base our predictions. Ho et al. (2008) approximate the NVSS galaxy angular overdensity as

$$\delta_{NVSS}(\hat{\mathbf{n}}) = \frac{1}{A} \int dr\, r^2 \bar{n}_g(r) b(r)\, \delta(r, \hat{\mathbf{n}}) =$$

$$\int dz\, f_{NVSS}(z)\, \delta(z, \hat{\mathbf{n}}), \qquad (28)$$

where $\delta(r, \hat{\mathbf{n}})$ corresponds to the dark matter density contrast, $r$ is comoving distance, $\bar{n}_g$ is the source average number density, $b(r)$ the source clustering bias, $A$ is a normalization factor and $f_{NVSS}(z)$ is fitted with a function of the form

$$f_{NVSS}(z) = \frac{\alpha^{\alpha+1}}{z_\star^{\alpha+1}\Gamma(\alpha)} b_{eff} z^\alpha \exp(-\alpha z/z_\star). \qquad (29)$$

The comoving distance $r$ and the redshift are related via the Hubble function, $dr/dz = H(z)$. The free parameters of this function, namely $\alpha$, $z_\star$ and $b_{eff}$ were obtained after cross correlating NVSS with other extragalactic surveys. Their results were $\alpha = 1.18$, $z_\star = 0.79$ and $b_{eff} = 1.98$.

## 4. WMAP data

The *Wilkinson Microwave Anisotropy Probe* (WMAP) is a CMB satellite that has been measuring intensity and polarization anisotropies in the millimeter range since the second half of 2001. It covers the frequency range 23 – 94 GHz, and the angular resolution ranges from 0.81° up to 0.21°, (see Komatsu et al. (2009) for latest results). This experiment has provided three different data releases (after one [WMAP1], three [WMAP3] and five [WMAP5] years of observations). By combining maps from the five different frequency channels (K, Ka, V, Q and W) plus some external information collected from different frequency ranges, the WMAP team is able to produce *clean* CMB maps, where a non-cosmological contribution (mostly generated by the Milky Way) is subtracted. The emission from extragalactic point sources cannot be so accurately removed, and the use of point source masks becomes necessary. After the release of WMAP5 data, a catalog of 390 point sources, all of them at the Jy level, was produced, and a mask excising a circle of 0.6° radius around each source was also provided (Wright et al. 2009). Of course, this is not a complete point source catalog, and subsequently different groups have provided larger source catalogs, (e.g., López-Caniego et al. 2007).

In our analyses we shall use the KQ75 mask (which is close to the conservative Kp0 mask built in the first data release) plus the point source mask released in WMAP5. Eventually, some analyses will be repeated with a mask built upon the more complete point source catalog of López-Caniego et al. (2007). Regarding the CMB data, most analyses will be conducted with the clean maps of bands Q, V and W of WMAP5 data. However, for the sake of comparison, some of those analyses will be repeated with the corresponding clean maps of WMAP1 and WMAP3 releases. In the paper we shall display results for the V band, unless otherwise explicited. We found negligible changes when switching to the Q and W bands.

## 5. Cross-correlation results

In this section we present the results for our cross-correlation analyses for each of the statistical methods outlined in Section (2). We consider scenarios of increasing degree of complexity:

---

[4] We tried to fit the NVSS pseudo-power spectrum with different models of biases that decrease with redshift, but they all failed to reproduced the observed NVSS pseudo- power spectrum unless the mean redshift of the source population was severely decreased.

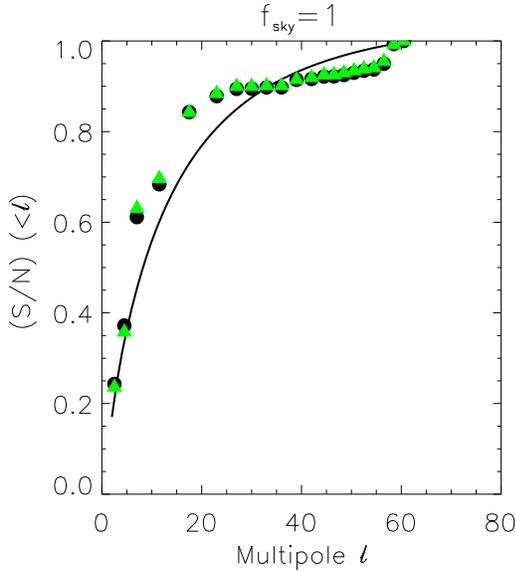
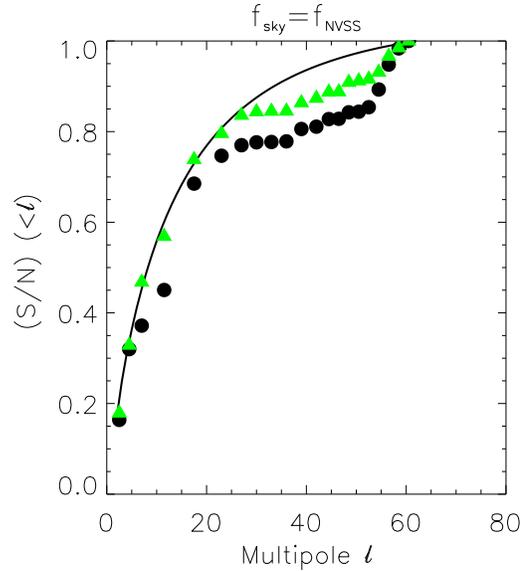

**Fig. 7.** Cumulative S/N below a given multipole $l$ for the MF (black circles) and ACPS (green triangles) approaches, ($f_{sky} = 1$).

**Fig. 9.** Same as in Figure (7), but after applying the WMAPV×NVSS effective mask.

we first apply our methods on an ideal simulation of the CMB, ISW and density fields under full sky coverage. Next we consider the effects of a NVSS-like sky mask on that same simulation, and finally we address the analysis of the real WMAP CMB and NVSS data.

### 5.1. The ideal case

We shall consider a CMB simulation under our WMAP5 cosmogony and a density field with the bias and redshift characterization corresponding to our reference model. The density field is gauged so that it has the correct degree of correlation with the ISW component of the simulated CMB field, as outlined in Section 2, (see also Paper I, Cabré et al. 2007; Barreiro et al. 2008).

#### 5.1.1. Full sky coverage

The left panel in Figure (6) shows the result of the implementation of the Fourier based MF and ACPS methods. Black circles and green triangles denote the estimates for the correlation coefficients $\alpha^{MF}$ and $\alpha^{ACPS}$, and the (black and green) solid lines display the corresponding 2-$\sigma$ confidence levels. Since in this case $f_{sky} = 1$, both methods should be identical, and hence provide very close estimates for both $\alpha$ and their error bars, (differences in the $\alpha$ estimates are slightly bigger due to innacuracies in the matrix inversion in the MF approach). Error bars in the $\alpha$ estimates were computed after 10,000 Monte Carlo (MC) simulations. These error bars can also be theoretically predicted in the case of the MF approach, and the agreement between both sets of error bars was of the order of ∼ 4% for $l < 30$. At higher multipoles the matrix inversion present in the theoretical prediction of $\sigma_{\alpha^{MF}}$ introduces some numerical noise, so we prefer to stick to the MC estimates, (see Paper I for details). Under these error bars, the significance of the cross-correlation according to the $\beta_j$ statistic defined in Section (2) is S/N ≃ 7 for both MF and ACPS methods, in full agreement with theoretical expectations.

In the middle panel of Figure (6) solid black circles display the sum of the diagonal terms of the $\mathbf{w}(\theta)$ array, which seem to be very close to the sum of all components of the same array (red triangles). This is to be expected, since for $f_{sky} = 1$, the impact of non-diagonal terms of $\mathbf{w}(\theta)$ must be necessarily small. The black solid lines display again the 2-$\sigma$ confidence level after 10,000 simulations, so according to these error bars at zero lag ($\theta = 0$) the S/N is close to 8. This significance must be interpreted cautiously, since those error bars were estimated after considering *only* the diagonal terms of $\mathbf{w}(\theta)$ and neglecting the correlation among different multipole bins for a fixed angle. I.e., at a given $\theta$, the correlation among the diagonal terms of $\mathbf{w}(\theta)$ was neglected, together with the errors of the off-diagonal terms. Hence, these confidence levels must be regarded as *optimistic*. The right panel compares the sum of the diagonal terms of $\mathbf{w}(\theta)$ (black filled circles) with the estimates of $w^R(\theta)$ under $N_{side} = 16$ (blue triangles) and $N_{side} = 64$ (green squares) HEALPix pixelizations. As expected for $f_{sky} = 1$, all estimates are very similar.

The S/N from the $\alpha$ estimates are shown in Figure (7). As before, black circles and green triangles correspond to the MF and ACPS approaches, respectively, whereas the solid black line displays the theoretical expectation. As mentioned above, practically half of the total S/N is found at $l < 10$, with only a small change (of ∼ 10 %) beyond $l > 30$. This confirms again that most of the information is restricted to the large angular scales.

#### 5.1.2. Partial sky coverage

Now we repeat the analyses under the joint mask of WMAPV and NVSS, and study the impact of the mask in the results. Regarding the Fourier based methods, the mask increases the coupling or correlation between neighbouring multipoles, but most importantly introduces some aliasing of power from the large to the small scales, as we shall see below. Since a smaller fraction of the sky is subject to be analysed, the S/N ratios in all methods decrease correspondingly. Results from the MF and ACPS methods are given in the left panel of Figure (8): since both methods perform slightly different if $f_{sky} < 1$, their $\alpha$ estimates and error bars do not coincide anymore. Also, the S/N has dropped in both cases (S/N ≃ 4, 5 for MF, ACPS, respectively) if compared to the full sky case (S/N ≃ 7). The middle panel also displays a smaller significance of the ACF at zero lag (S/N $\simeq$ 6), although non-diagonal terms in the $\mathbf{w}(\theta)$ array seem to still have little impact (the addition of diagonal terms (black circles) is very close to the addition of all terms of $\mathbf{w}(\theta)$, (red triangles)). The subtraction of the mean and dipole in the whole sphere (even though there are masked regions) is not equivalent to the monopole subtraction *outside* the mask (as it is done for $w^R(\theta)$), and for this reason the diagonal terms of $\mathbf{w}(\theta)$ do not coincide with the estimation of $w^R(\theta)$ under $N_{side} = 16$ (green squares) and $N_{side} = 64$ (blue triangles) pixelizations.

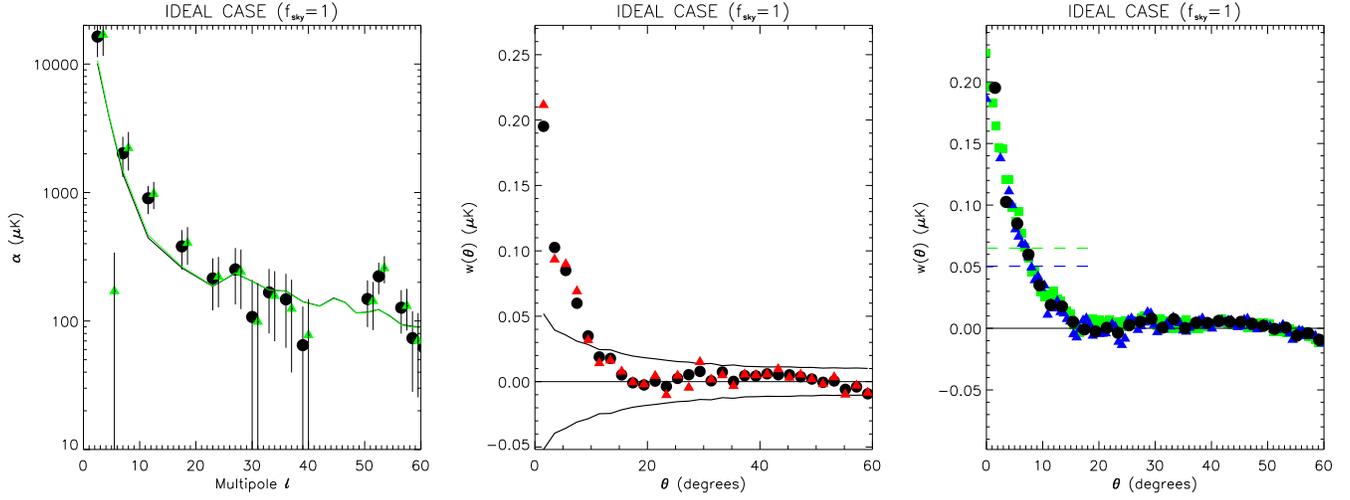

**Fig. 6.** *Left panel:* Cross-correlation coefficient estimates ($\alpha$-s) versus multipole for the MF (black circles) and ACPS (green triangles) methods ($f_{sky} = 1$). Solid lines depict the 2-$\sigma$ levels for each method (which coincide since $f_{sky} = 1$). *Middle panel:* Sum of diagonal (black circles) and all (red triangles) terms of the $\mathbf{w}(\theta)$ array. Solid lines display 2-$\sigma$ level contours. *Right panel:* Three different estimates of the ACF: diagonal terms of $\mathbf{w}(\theta)$ (black circles), and $w^R(\theta)$ estimates under $N_{side} = 16$ (blue triangles) and $N_{side} = 64$ (green squares). Blue and green dashed lines provide the 2–$\sigma$ confidence level for $w^R(\theta = 0)$ after 100,000 MC simulations, for $N_{side} = 16, 64$ pixelizations, respectively.

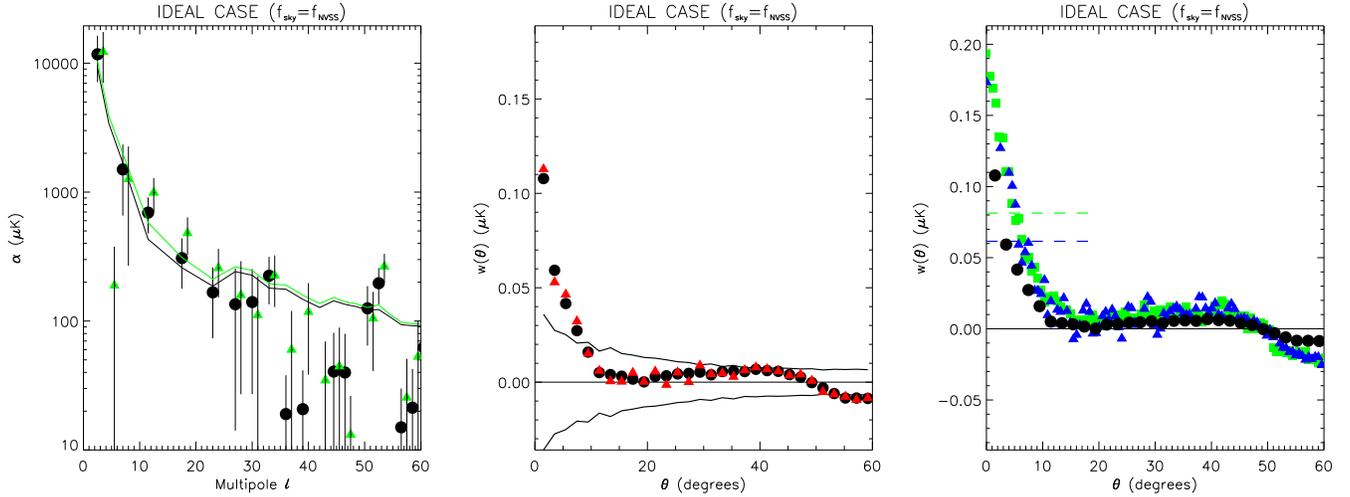

**Fig. 8.** Same as in Figure (6), but after applying the WMAPV×NVSS effective mask.

The impact of aliasing is well visible in Figure (9), in particular for the MF method: due to the mask, part of the low $l$ power is transferred into small angular scales, and between 10% and 20% of the total S/N is shifted to $l > 40$. However, we still have the case where practically half of the total S/N is located at $l < 10$, and roughly 80%-90% at $l < 40$.

### 5.2. Real data case

The two ideal cases shown above seem to be pretty far from the real WMAP5-NVSS cross correlation, as Figures (10,11) show. The top row in Figure (10) shows $\mathbf{w}(\theta)$ for each of the three NVSS templates considered, (black circles correspond to diagonal terms, red triangles to the sum of all terms). The solid lines display the 2-$\sigma$ level computed after 10,000 MC realizations. The amplitude at the origin is positive in the three considered cases, but below the level of half a sigma. This is compatible with the bottom row, where the cross correlation coefficients $\alpha$ are shown for different multipole bins (as above black circles and green triangles correspond to the MF and ACPS approaches, respectively). In the multipole range $l \in [2, 60]$, there is no single point above the 2-$\sigma$ level: only from $l \simeq 10$ up to $l \simeq 23$ there are three consecutive points above the zero level, which, if isolated from all other points, could give rise to some significance at the 2-$\sigma$ level. However, when considering other multipole ranges, or the entire interval $l \in [2, 60]$, the considerably symmetric scatter around zero prompts to a very low significance for both $\beta_{N_b}$ and $\mathbf{w}(\theta)$, as it is the case.

| % error | $N_{side} = 16$ | $N_{side}=64$ |
|---|---|---|
| $S_\nu > 2.5$ mJy | 1.5 | 2.3 |
| $S_\nu > 30$ mJy | 2.9 | 2.4 |
| $S_\nu > 60$ mJy | 0.7 | 2.8 |

**Table 1.** Relative mismatch (in %) of the zero lag errors of the ACF ($w^R(\theta)$) estimates after 500 MC realizations when comparing them to the outcome of 100,000 MC simulations (evaluated only at $\theta = 0$).

We next compare the estimates of $\mathbf{w}(\theta)$ with the estimates of $w^R(\theta)$ under $N_{side}$ =16, 64 HEALPix pixelizations (see Figure (11)). Since the two approaches for removing the *all sky* monopole and dipole (in one case) and monopole *outside the mask* (in the other) are different, the estimates of the ACF-s and their error bars do not coincide. The solid green lines display the 2-$\sigma$ contour levels for the $w^R(\theta)$ under $N_{side} = 64$ computed after 500 MC simulations. We checked that the amplitude of the error bars at zero lag ($\theta = 0$) converged to the average value of 100,000 MC simulations conducted at that same $\theta$ point, as we shall discuss later. For the thresholds at $S_\nu > 30$, 60 mJy it is clear that the ACF estimate for $N_{side} = 64$ points to some significant cross-correlation above the level of 2-$\sigma$ (green squares). This significance drops slightly (at the level of 2-$\sigma$) for the $N_{side} = 16$ estimate (blue triangles), and in both pixelizations the excess decreases by more than 50% at a distance $\theta \simeq 4°$.

## 6. Discussion

### 6.1. Convergence of the ACF errors and other consistency tests

When computing the errors associated to the estimates of the ACF-s, we have followed two different approaches. Given the cost (in CPU time) to compute each estimate for $w^R(\theta)$ for all angles, we have run *only* 500 MC simulations to estimate the error bars shown in Figure (11). Nevertheless, it is much faster to estimate the ACF at zero lag ($\theta = 0$) separation, since it involves the generation of a random CMB map and a mere multiplication of such map with the (fixed) density template (modulo some trivial manipulations, like the removal of the monopole outside the effective mask). In Table (1) we provide the relative difference (in %) between the ACF error estimates from the 500 MC simulations of $w^R(\theta)$ in one side, and the error estimates from 100,000 MC simulations in the other, always at zero lag ($\theta = 0$). The convergence after 500 realizations was better for the $N_{side} = 16$ case, for which only errors in the first angular bin (where $w^R(\theta)$ was evaluated) were considered when comparing with the average of the 100,000 MC simulations. For $N_{side} = 64$, we needed to average the errors in the first three angular bins in order to reach good agreement with the outcome of 100,000 MC simulations. I.e., when looking at the error bars in Figure (11) (green solid lines), one should take the average the amplitude of the errors in the first three bins at low $\theta$. Having this present, the agreement was better than 3%, as Table (1) shows.

We also computed the errors at zero lag for the ideal case ($f_{sky} = 1$) considered in Section (5.1), and found that the errors from 100,000 MC simulations coincided (within 5%) with the theoretical estimates given by the denominator of Equation (19):

$$\Delta[w(\theta = 0)] = \frac{1}{(4\pi)^2} \sum_{l=2}^{l_{max}} (2l+1) C_l^g C_l^{LSS} (W_l^{PSF})^4. \quad (30)$$

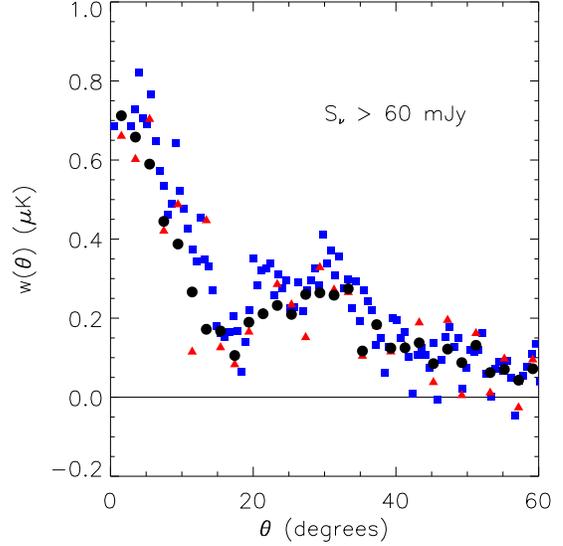

**Fig. 12.** Comparison of the ACF estimate $w^R(\theta)$ with the estimate from the array $\mathbf{w}(\theta)$. The ACF $w^R(\theta)$ is computed between the WMAP5 V band temperature map and the NVSS template ($S_\nu > 60$ mJy) *where monopole and dipole in the whole sphere have been subtracted*, for a HEALPix $N_{side} = 16$ pixelization, and is displayed by blue squares. The trace and sum of all terms of $\mathbf{w}(\theta)$ computed on the same maps are displayed by red triangles and black circles, respectively.

Note that this equation is different from Equation (17) of Vielva et al. (2006), since we consider our density template to be *fixed* and that the errors come exclusively from the signal generated at the last scattering surface. The fact that the errors of the 500 realizations of $w^R(\theta)$ coincide with the error estimation at zero lag after 100,000 simulations, and that this error estimation at zero lag coincides with the theory prediction in the ideal case points to the correctness of the error bars from $w^R(\theta)$ estimates shown in Figure (11). Furthermore, we computed the real space ACF $w^R(\theta)$ on the full sphere between WMAP maps and our three NVSS density templates *after removing the monopole and the dipole in the entire sky*. This estimate of $w^R(\theta)$ should coincide with the sum of all terms of the array $\mathbf{w}(\theta)$, and the comparison of both can be seen in Figure (12). Let us remark that the two pieces of software (one computing $w^R(\theta)$, the other $\mathbf{w}(\theta)$) are *independent*, and operate in real and Fourier space, respectively. Nevertheless they yield the same result.

### 6.2. Comparison with previous results

#### 6.2.1. In Fourier/wavelet space

When comparing WMAP to NVSS data, there exist several claims for ISW detection in wavelet space, namely Vielva et al. (2006); McEwen et al. (2007); Pietrobon et al. (2006). Those works are compatible to each other, since they find an excess of correlation at the 3.5 – 4 $\sigma$ level at scales between 2°and 10°. However, they are all incompatible with the more recent study of Ho et al. (2008), which claims a $\sim$ 3–$\sigma$ detection at $l \simeq 6$, i.e., scales considerably larger than 10°. It is worth to note that, in that analysis, large angular scales (below $l < 10$) are regarded as dominated by *spurious power*, and they are ignored when interpreting the NVSS auto-power spectrum. However, they are used when interpreting the WMAP–NVSS cross power spectrum, since they precisely host most of the correlation excess.

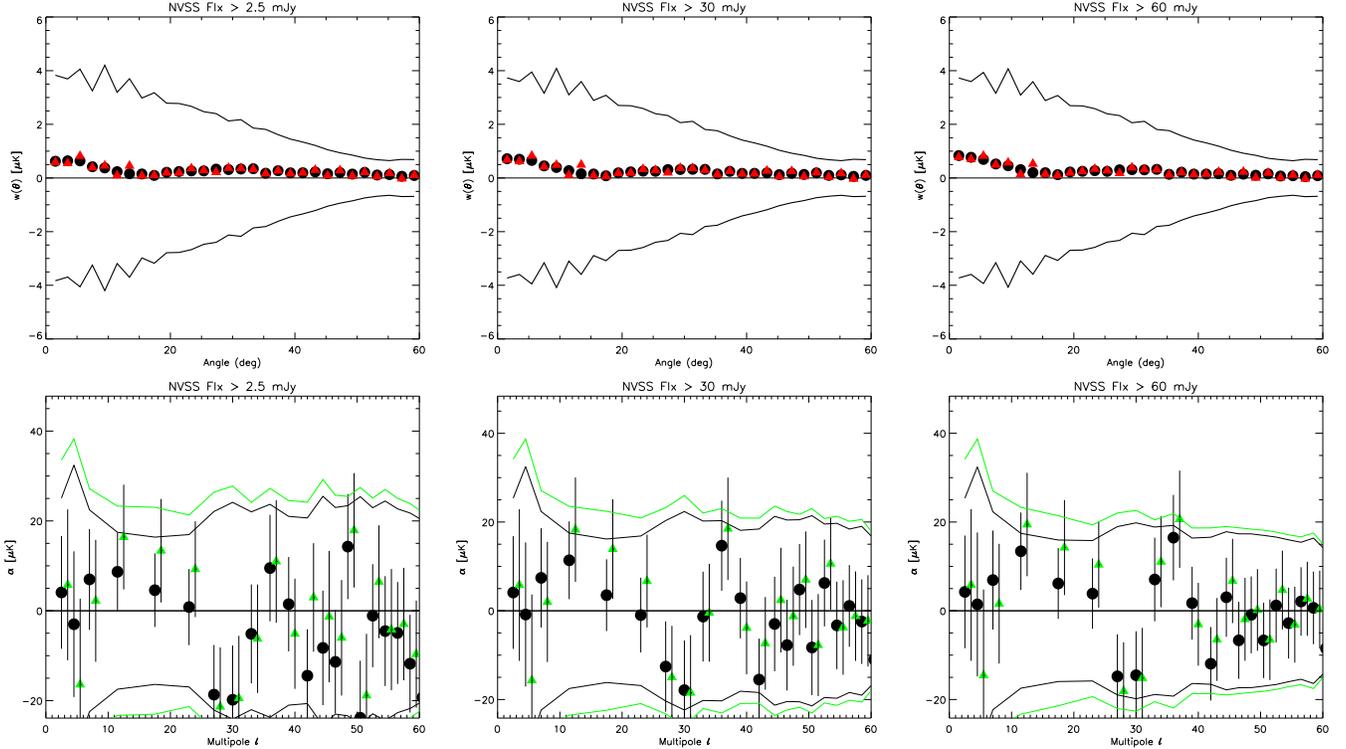

**Fig. 10.** *Top row:* Estimates of the ACF between WMAP5 data and NVSS radio galaxies under a flux threshold of 2.5 (left), 30 (center) and 60 (right) mJy. *Bottom row:* Corresponding Fourier cross-correlation coefficients ($\alpha_i$-s) estimates for each of the three cases considered in the top row.

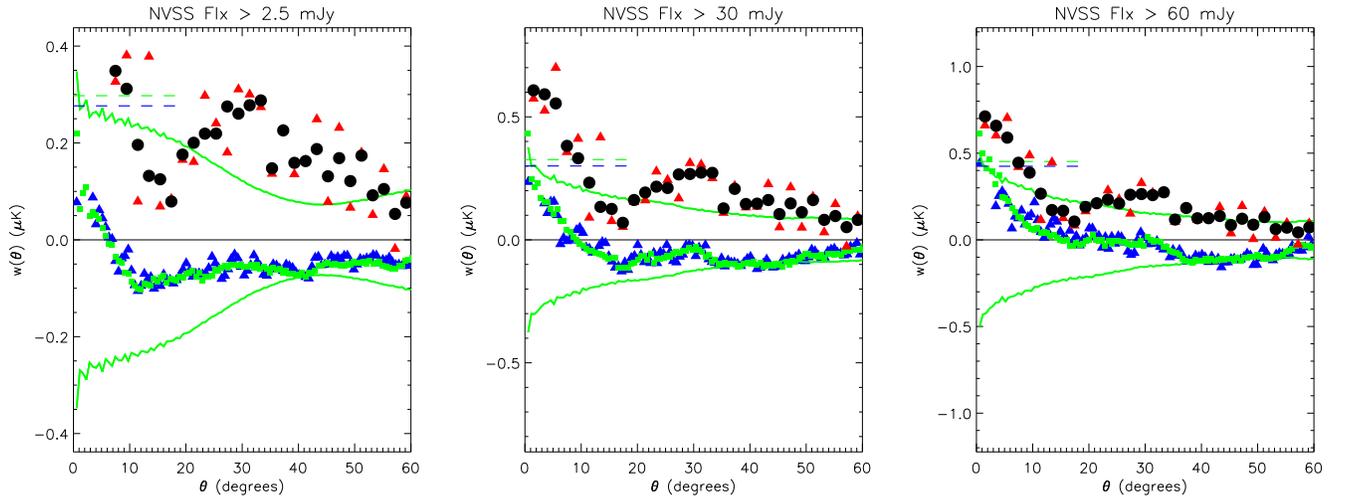

**Fig. 11.** ACF from WMAP5 and NVSS data for radio sources brighter than 2.5 (left), 30 (center) and 60 (right) mJy. The color and symbol coding is as in previous plots: diagonal (all) terms of the array $\mathbf{w}(\theta)$ are given by black circles (red triangles). Estimates of $w^R(\theta)$ at $N_{side} = 16, 64$ pixelizations are given by green squares and blue triangles, respectively. The solid green lines depict the 2-$\sigma$ level for the $w^R(\theta)$ ($N_{side} = 64$) estimate. Green and blue dashed lines show the 2–$\sigma$ level at zero lag ($\theta = 0$) for $N_{side} = 64$ and $N_{side} = 16$ pixelizations, respectively. These confidence levels do not apply for $\mathbf{w}(\theta)$ estimates (black circles and red triangles).

The method used by Ho et al. (2008) is close to a cross power spectrum in multipole space, but optimized to yield minimum variance, (see Padmanabhan et al. 2005). Let us note as well that Ho et al. (2008) find a 2–$\sigma$ result in the multipole bin centered around $l \simeq 20$.

Our results in multipole space are in partial disagreement with those of Ho et al. (2008), since for $l < 10$ we find no hint of positive cross-correlation, and this translates in a rather flat ACF at large angles, (top panels of Figure (10)). As mentioned above, our MF and ACPS approaches find some hint of positive cross-correlation at scales $l \in [10, 25]$, but at low significance level (~ 2–$\sigma$ in that restricted multipole range. This is in better agreement with Ho et al. (2008), who obtain a ~ 2-$\sigma$ detection at $l \simeq 20$). This apparent excess could marginally correspond to the signal found in wavelet space, although those claims are based upon slightly smaller angular scales and much higher statistical significance. In summary, our results point to some low significance evidence (~ 2–$\sigma$) of cross-correlation in the multipole range $l \in [10, 25]$, in agreement with Ho et al. (2008). Methods based upon wavelets find a much more significant sig-nal at slightly smaller angular scales, and those results and ours fail to see any signal below $l < 10$, in clear disagreement with Ho et al. (2008). In our case, when integrating in the whole multipole range $l \in [2, 60]$, we fail to detect any significant cross-correlation between NVSS and WMAP data.

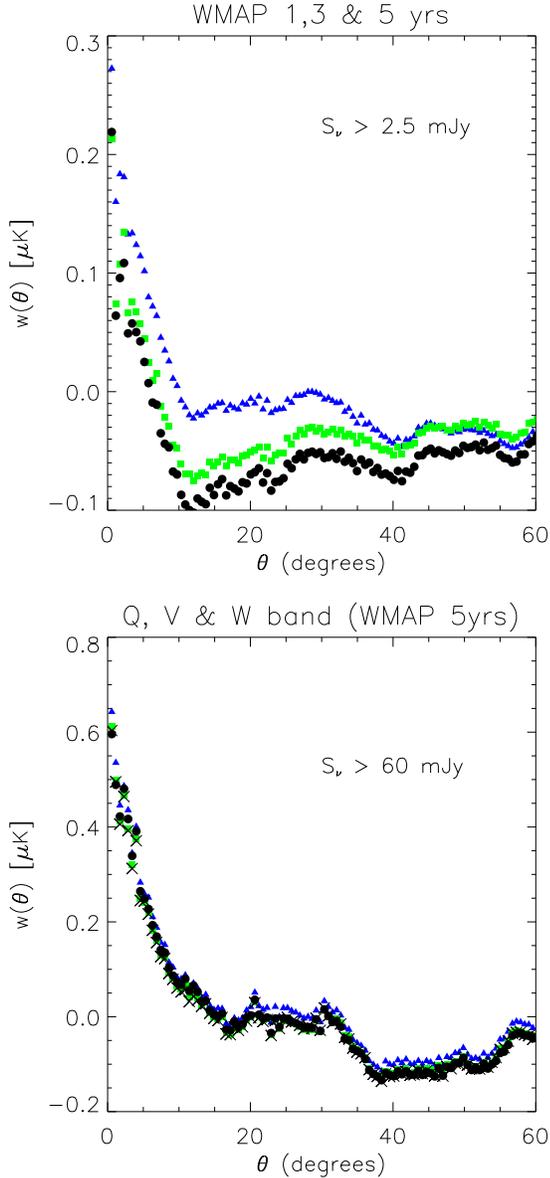

**Fig. 13.** *Top panel:* Estimates of the ACF $w^R(\theta)$ for the lowest threshold considered in NVSS ($S_\nu > 2.5$ mJy) and for WMAP first (blue triangles), three (green squares) and five (black circles) year data. *Bottom panel:* Estimates of the ACF $w^R(\theta)$ for the highest threshold considered in NVSS ($S_\nu > 60$ mJy), and the cleaned Q (blue triangles), V (green squares) and W (filled circles) WMAP5 temperature maps. Crosses denote the result for the V band, but under the point source mask of López-Caniego et al. (2007).

### 6.2.2. In Real Space

In real space the comparison improves significantly. The shape of the ACF $w^R(\theta)$ computed after using WMAP1 data (blue triangles in the top panel of Figure (13)) is in good agreement with the ACF found by Boughn & Crittenden (2004); Nolta et al. (2004); Vielva et al. (2006); Raccanelli et al. (2008). In all cases, the ACF approach zero at $\theta \sim 10°$, and remains rather flat for bigger angles. In our analysis, and in those of Boughn & Crittenden (2004); Vielva et al. (2006); Raccanelli et al. (2008) the ACF drops steadily at low $\theta$'s, so $w(\theta = 0°)/w(\theta = 3°) \simeq 2$, whereas in that of Nolta et al. (2004) the ACF remains roughly constant, ($w(\theta = 0°)/w(\theta = 3°) \simeq 1$). The significance at zero lag is, in our case, at the level of 2–$\sigma$, which is slightly below the 2.5–$\sigma$ found by Boughn & Crittenden (2004) and the 3–$\sigma$ found by Vielva et al. (2006); Raccanelli et al. (2008), but quite close to Nolta et al. (2004). Overall, results are quite similar, with small differences which are likely caused by distinct approaches in the processing of the source catalogs, their projection onto a HEALPix format, the construction of the point source masks and the actual implementation of the cross-correlation algorithm.

The impact of diffuse galactic foregrounds is assessed as follows. The top panel of Figure (13) shows the dependence of the ACF $w^R(\theta)$ when the same NVSS template ($S_\nu > 2.5$ mJy) is cross-correlated with WMAP1 (blue triangles), WMAP3 (green squares) and WMAP5 (black circles) data. In all cases we used the *clean* versions of the V-band maps. As the contaminant subtraction improves, the shape of the ACF changes consequently, but in all cases there is a sudden drop of the ACF from zero lag ($\theta = 0$) to non zero separation angle ($\theta > 0$). Although significant changes should not be discarded for any future WMAP data release, the comparison between the second and third (and last to date) data releases prompted modifications in the ACF that are of minor relevance, (i.e., the difference between green squares and black circles in the top panel of Figure (13) does not have any impact in our discussion).

As Figure (11) shows, the zero lag statistical significance of the ACF increases up to $\sim 3$–$\sigma$ as higher radio flux thresholds are applied. Nevertheless, we must keep in mind that the error bars of our ACF estimates must be regarded as *optimistic*, since they ignore the part of the ISW component that is not correlated to the galaxy survey (small changes are to be expected, though).

### 6.3. Evidence for ISW?

Our analyses fail to yield conclusive evidence of ISW in the NVSS - WMAP cross correlation. The estimates of $w^R(\theta)$ provide an excess of correlation at zero lag that is dependent of the flux threshold, but the clustering of radio sources is independent of flux level (as shown in Figure 4) and so the ISW – correlated signal should be. There should be no significant difference (at most 10%) at zero lag for the $w^R(\theta)$ estimates between the $N_{side} = 16$ and the $N_{side} = 64$ pixelizations, as found in Figure (8), but in practice the zero lag amplitude drops by $\sim 30\%$–$50\%$ in the low resolution case. This is all expected after looking at the ACF $\mathbf{w}(\theta)$ estimates in the top row of Figure (10): when the ACF computation is restricted to big angular scales ($l < 60$), there is no apparent correlation excess.

All these tests seem to point to source emission as responsible for the correlation excess found with the $w^R(\theta)$ estimates. These show, however, little frequency dependence, as displayed in the bottom panel of Figure (13), and this could be used as an argument against intrinsic radio source emission. The cleaning algorithms are designed to remove the *frequency dependent* signal in the CMB temperature maps, but it is not clear how such procedures subtract the constant level that does not change from frequency to frequency (especially if no template for emission is available for NVSS sources). In these circumstances, algorithms removing foreground emission are likely to subtract the frequency dependent part of the source signal, but may leave a constant DC level at the sources' position. This seems to be the most straightforward interpretation of the flux threshold dependent correlation excess found in the ACF $w^R(\theta)$ estimates. It is known that many of those radio sources are also bright in the infrared (Helou et al. 1985; Seymour et al. 2009), and there is also evidence for a turning in the spectral index of dim radio sources at high frequencies, (Lin et al. 2009).

We lack high significance evidence for the ISW, but radio source emission is a clear foreground in WMAP temperature maps on the small angular scales. Even after masking the few hundred brightest sources, there must still be contribution from the dimmer population, and this has not been formally addressed so far in ISW studies. Regardless of the presence of ISW signal in WMAP scans, our analyses seem to provide evidence for residual radio source emission *outside* the usual foreground/point source masks, which should be considered when addressing ISW cross correlation analyses. If this interpretation is correct, then most of the source emission must be coming from a source population below the flux threshold imposed by the point source mask: the use of the more aggressive mask of López-Caniego et al. (2007), which covers $\sim 330$ extra square degrees with relatively bright radio sources, introduces no significant difference (see crosses in bottom panel of Figure (13)).

The estimate of the ACF based upon the $\mathbf{w}(\theta)$ computation links our Fourier space cross correlation analyses (MF and ACPS) with the $w^R(\theta)$ estimation in real space. If the excess of zero lag correlation found with the $w^R(\theta)$ is real and corresponds to radio emission present in WMAP temperature maps, then the agreement between the ACF $w^R(\theta)$ and $\mathbf{w}(\theta)$ estimates shown in Figure (12) demonstrates that the Fourier analyses displayed in the bottom row of Figure (10) actually constitute the Fourier counterpart of the ACF $w^R(\theta)$ results given in Figure (11), (although restricted to large angular scales). These analyses find no evidence for significant cross correlation in the angular range where the ISW – density coupling is supposed to arise. Only in the multipole range $l \in [10, 25]$ there is some marginal, low significance ($\sim 2-\sigma$) signal which could correspond to the ISW detection claims of Vielva et al. (2006); McEwen et al. (2007); Pietrobon et al. (2006). Were such signal really corresponding to the ISW effect, the puzzle would still consist in explaining why there is no evidence for ISW on the very large scales ($l < 10$), where theoretically most of the S/N should be found. The results of Ho et al. (2008), although incompatible with the rest in this large angular range, are in better agreement with theoretical predictions for the angular scales of the S/N. Nevertheless, their most significant detection claim is located precisely in the low multipole range ($l < 10$) where they find dominant spurious power and which is discarded when characterizing the NVSS auto-power spectrum.

This apparent big angle contamination cannot be explained in terms of declination dependent sensitivity of the NVSS survey: the low multipole/big angle clustering properties are identical for NVSS sources above 2.5, 30 and 60 mJy, and in the last two cases there is no declination dependence of the source angular number density. Furthermore, sources above 30 and 60 mJy are very clearly detected (S/N $\sim 30$–$60$) and it is not easy to find any contaminant that may bias the clustering of so clearly detected sources. Blake et al. (2004) find a similar level for NVSS big angle clustering, and conclude that most of the NVSS angular power spectrum must be generated at low redshifts, $z < 0.1$. This would explain the lack of evidence for ISW in our analyses, but contradicts the model of Dunlop & Peacock (1990) and the thorough analysis of Ho et al. (2008) for the redshift distribution of NVSS sources. This is indeed a key issue when interpreting the WMAP – NVSS cross-correlation. According to our results,

*(i)* either the NVSS source population is located at low redshift (so that it cannot properly probe the ISW), or *(ii)* the source population is located at high redshift but shows an abnormal power spectrum on the big scales and does not probe properly the spatial distribution of the large scale gravitational potential wells, or *(iii)* the source population is located at high redshift, it shows an abnormal power spectrum on the big scales, it may correctly trace the spatial distribution of potential wells, but there is no ISW component in WMAP data. We understand that none of this three scenarios can be easily accommodated.

## 7. Conclusions

In this work we analyze the evidence for ISW in the cross correlation of CMB data and the NVSS radio galaxy catalog. For that, we have conducted a two point function cross-correlation analysis between the NVSS galaxy survey and WMAP CMB data both in Fourier and in real space. We have implemented them in two different, independent pieces of software which, when projected both onto real space, agree with each other. The real space two point estimator is more sensitive to the small angular scales, whereas the Fourier cross correlation estimator is able to isolate the contribution from different angular/multipole ranges, and enables a more direct and detailed comparison with theoretical predictions.

We have studied the NVSS radio survey, and constructed density templates after applying three different flux thresholds at 2.5, 30 and 60 mJy. The known systematic of the source angular number density with respect to ecliptic declination disappears when only bright sources ($S_\nu > 30$, 60 mJy) are selected. Moreover, when computing the angular power spectrum of each of those density templates, we reproduce the expected level for the Poisson (or shot noise) term at high multipoles (or small angles), and find that the clustering of the NVSS sources (present at low multipoles or big angles) does not depend on the flux threshold. This clustering is however incompatible with the model built by Ho et al. (2008) on the redshift dependence of NVSS source bias and density, at least on the very large scales.

Nevertheless, we make use of this model to make predictions for the detectability of the ISW via a cross correlation between CMB and NVSS data under a WMAP5 cosmogony. The ideal total S/N for this cross correlation should be close to 7, which under the WMAP5 x NVSS effective mask should decrease to S/N$\sim 5$. In both cases, half of the total S/N should be arising at low multipoles ($l < 10$), and 80%-90% of it at $l < 40$. These idealized simulations allowed us testing and calibrating our cross correlation algorithms, which were next applied onto real NVSS and WMAP data.

In real space, our analyses found statistical evidence (from $2-\sigma$ to $3-\sigma$) for an excess of correlation at zero lag of the angular correlation function. This excess is more significant for higher flux thresholds applied on the NVSS radio source population, drops to $\sim 50\%$ of its value at an angular distance of $\theta \simeq 4°$, and decreases as well for a 30%–40% when increasing the pixel size from $\theta_{pix} \simeq 1°$ ($N_{side} = 64$) to $\theta_{pix} \simeq 4°$ ($N_{side} = 16$). Such strong dependence on the separation angle $\theta$ is not predicted by theory nor reproduced by our simulations of ISW – density cross correlations. When comparing to previous results, we find rough agreement in the shape and significance of the angular correlation function, but our tests on the dependence on flux threshold and pixel size tend to suggest the presence of residual point source emission in WMAP cleaned maps.

If we restrict our analyses to large angular scales ($l < 60$), we find no evidence for correlation excess, in real nor in Fourier space, independently of the flux threshold applied on NVSS sources. Only in the multipole range $l \in [10, 25]$ (or $\theta \in [10°, 20°]$) we find some marginal excess (at the 2–$\sigma$ level) which could be in relation with previous ISW detection claims from wavelet-based analyses. But the lack of signal at the very large angular scales ($l < 10$ or $\theta > 20°$) is in clear contradiction with theoretical predictions, which state that most of the S/N of the cross correlation analysis should be generated at those angles. This discrepancy in the S/N arises precisely at the angular range where the theoretical description of NVSS auto-power spectrum fails to reproduce the actual clustering of radio sources. Our results suggest that either current models placing NVSS sources at high redshift are incorrect, or NVSS sources are indeed at high redshift but show abnormal clustering at low multipoles. Were this the case, then either those sources do not trace correctly the large scale gravitational fields, or the ISW component in WMAP data is for some reason obscured or absent.

Since none of those conclusions is satisfactory, we are conducting similar analyses on galaxy surveys whose bias and redshift distribution is more accurately characterized than for NVSS. Nevertheless, this involves working with shallower, smaller area surveys for which the expected S/N from ISW – density cross correlation is not as high as for the NVSS under our reference model. Currently, analyses based upon the Sloan Digital Survey[5] (SDSS) are underway.

*Acknowledgements.* I am grateful to Ricardo Génova–Santos, Roderik Overzier and Fernando Atrio–Barandela for useful discussions. I also acknowledge enriching interaction with Patri Vielva, Björn Schäfer, Marian Douspis, Nabila Aghanim, Jordi Miralda–Escudé, Claudia Scóccola, Francesco Shankar, Raúl Angulo, Eric Switzer, D.N.Spergel and K.Smith. I acknowledge the use of the HEALPix (Górski et al. 2005) package and the LAMBDA[6] data base.

---

[5] SDSS URL site: http://www.sdss.org/
[6] http://lambda.gsfc.nasa.gov